\documentclass[11pt]{article}
\usepackage[latin1]{inputenc}
\usepackage[english]{babel}
\usepackage[namelimits]{amsmath}
\usepackage{amssymb}
\usepackage{amsmath}
\usepackage{amsthm}

\begin{document}
\title{Entropy, temperature and internal energy of trapped gravitons and corrections to the Black Hole entropy}
\author{
Stefano Viaggiu,\\
Dipartimento di Matematica,
Universit\`a di Roma ``Tor Vergata'',\\
Via della Ricerca Scientifica, 1, I-00133 Roma, Italy.\\
E-mail: {\tt viaggiu@axp.mat.uniroma2.it}}
\date{\today}\maketitle
\begin{abstract}
\noindent In this paper we study the proposal present in \cite{1} concerning the statistical description of 
trapped gravitons and applied to derive the semi-classical black hole
(BH) entropy $S_{BH}$.
We study the possible configurations depending on 
physically reasonable expressions for the internal energy $U$. In particular, we show that expressions for 
$U\sim R^k, k\geq 1$, with $R$ the radius of the confining spherical box, can have a semi-classical description, while behaviors with
$k<1$ derive from thermodynamic or quantum fluctuations.
There, by taking a suitable physically motivated expression for $U(R)$, we obtain the well known logarithmic corrections to the 
BH entropy, with the usual behaviors present in the literature of BH entropy. Moreover, a phase transition 
emerges with a positive specific heat $C$ at Planckian lengths instead of the usual negative one
at non-Planckian scales, in agreement with results present in the literature. Finally, we show that evaporation stops at a radius 
$R$ of the order of the Planck length.

\end{abstract}
%PACS Nos.: 05.20.-y, 04.30.-w, 04.70.-s, 04.60.-m

\section{Introduction}
Since of the fundamental discovery that \cite{2,3} BH emit a thermal radiation and as a consequence they have a non-vanishing
entropy $S_{BH}$ with $S_{BH}=K_B A_h/L_P^2$ (with $K_B$ the Boltzmann constant, $L_P$ the Planck length and 
$A_h$ the area of the event horizon), many attempts
(see \cite{4,5,6,7,8,9,10,11,11b,12,13,14,15} to cite someone) appeared in the literature in order to explain where are stored the microscopic
degrees of freedom leading to the BH formula, i.e. the statistical understanding of the BH entropy. Generally, the
entropy formula derivation is related to the unproven used theory (loop, string), while in \cite{12,13,14,15} the modes are given by 
considering  the quasi-normal BHs frequencies. 
Moreover, in the literature
the BH entropy is derived by considerations outside the event horizon. However, it is physically reasonable and 
intriguing to ask what happens inside a BH. No experimenter can escape from the spacelike Schwarzschild singularity, so he/her can only measure for a short time. 
Nevertheless, the study from the point of view inside the event horizon, although looks like a speculative hypothesis, it
is physically interesting and intriguing. In fact, it is a possible response to the question regarding the nature of the degree of
freedom leading to the BH entropy\footnote{For a nice review see \cite{Page} and references therein.}. 
This perspective has been addressed in \cite{1}. There, an approximate formula is derived 
describing the discrete spectrum of trapped gravitons inside a spherical box and applied in order to obtain the well known BH
expression from a statistical point of view. The internal degrees of freedom are provided by $N$ harmonic oscillators composed of 
gravitons. Why gravitons? First of all, we expect that inside a BH only radiation can escape from the spacelike singularity.
The BH solution is a vacuum solution of Einstein's equations, so we expect that if something can survive inside a 
BH, then it should verify that its energy-momentum tensor $T_{\mu\nu}$ is identically vanishing, i.e. gravitons. 

In this paper we continue the investigation present in \cite{1}. In particular, we focus our attention on the possible 
viable (classical and quantum) expressions for the internal energy $U(R)$ and the related expression for
the temperature $T$ and the entropy.
For the BH case we test our approximate formula for the spectrum of trapped gravitons against theoretical predictions by showing a good agreement. We show that the logarithmic corrections in the IR limit can be easily obtained in our context. Also the
Planckian limit is investigated, where non commutative effects come into action.
Finally, we attempt to generalize our treatment to a Kerr-Newman BH.

In section 2 we shortly present the formalism and the general formulas. In section 3 we study 
'classical' cases with viable expressions for the internal energy $U$. In section 4 we analyze the BH case. In section 5 we study the way to obtain corrections to the semi-classical BH entropy, while in section 6 we derive some interesting consequences due to the corrections obtained.
Finally, section  7 is devoted to a discussion of the results obtained together with possible generalizations and outlooks.

\section{Preliminaries}

To obtain a suitable formula for the spectrum of trapped gravitons, in \cite{1} we considered a gravitational wave traveling in
a vacuum Minkowski spacetime $g_{ik}^{(0)}$ together with the perturbation $h_{ik}$ representing a gravitational wave.
As well known \cite{16}, for a gravitational wave in a Minkowski spacetime, the perturbed and linearized Einstein's equations reduce to a master equation in
terms of the axial (Regge-Wheeler) $Z_{\ell m}^{(a)}(r)$ and polar (Zerilli) $Z_{\ell m}^{(p)}(r)$ functions given by
\begin{equation}
Z_{\ell m,r,r}^{(p,a)}+{\omega}^2 Z_{\ell m}^{(p,a)}=\frac{\ell(\ell+1)}{r^2}Z_{\ell m}^{(p,a)},
\label{1}
\end{equation}
where $\{\ell m\}$ are the integer Legendre indices and $\omega$ the frequency of the traveling wave.
The solution of the (\ref{1}) for $Z_{\ell m}^{(a)}(r)$ is proportional to the Bessel functions $j_{\ell}(kr)$ with
\begin{equation}
j_{\ell}(kr)\simeq\frac{1}{kr}\cos\left[kr-\frac{(\ell+1)\pi}{2}\right],
\label{2}
\end{equation}
where $k=\omega/c$.
After imposing the Dirichlet boundary condition\footnote{See \cite{1} for more details.} 
$Z_{\ell m}^{(p,a)}(R)=0$, being $R$ the areal radius of the confining spherical box,
we obtain the following approximate formula for the angular frequency ${\omega}_{\ell n}$ of the trapped gravitons:
\begin{equation}
{\omega}_{\ell n}\simeq\frac{c}{2R}\left(2+\ell+2n\right)\pi,\,\;\ell\geq 2,\;\;n\in\mathbb{N}.
\label{3}
\end{equation}
Formula (\ref{3}) is a good approximation for the zeros of the Bessel function. 
In particular, the error of the first zero is of the order of $4\%$ with the error rapidly decreasing for the other zeros.
In section 4 we show that this is the case by comparing the parameters giving the BH entropy derived from the (\ref{3})
with the ones from exact formulas. In \cite{1} we promoted $n$ to a quantum number, while the Legendre index $\ell$ has been 
identified as a 'species number' depicting the species of gravitons present inside the box, i.e. $\ell=2$ quadrupolar,
$\ell=3$ sextupolar..., without promoting the azimuthal index $m$ to a quantum number. In section \cite{4} we show that this is 
the correct interpretation. Also note that, according to the finding present in \cite{12,13,14,15}, since for the
(\ref{3}) we have $\omega R\sim n$ for large values of the quantum number\footnote{For any fixed $\ell$, for $n>>\ell$.} 
$n$, the spectrum
(\ref{3}) is in agreement with the Bohr's correspondence principle, i.e. transition 
frequencies behave as the ones of a classical oscillator for large $n>>1$.

To start with, we can calculate the partition function 
$Z_T=Z_g^N$, with $N$ harmonic oscillators with
\begin{equation}
Z_g=\sum_{\ell=2}^{\infty}\sum_{n=0}^{\infty} e^{-\beta\hbar{\omega}_{\ell n}}=
\frac{e^{-\left(\frac{2c\pi\beta\hbar}{R}\right)}}
{\left[1-e^{-\left(\frac{c\pi\beta\hbar}{2R}\right)}\right]}
\frac{1}{\left[1-e^{-\left(\frac{c\pi\beta\hbar}{R}\right)}\right]},
\label{4}
\end{equation}
where $\beta=1/(K_B T)$.
For the entropy $S$, the internal energy $U$ and the pressure $P$, thanks to the (\ref{4}) we obtain
\begin{eqnarray}
& & S=-N K_B\left[\ln\left(1-e^{-\frac{X}{2}}\right)+\ln\left(1-e^{-X}\right)\right]+\label{5}\\
& & +\frac{c\pi\hbar N e^{-\frac{X}{2}}\left[1+3 e^{-\frac{X}{2}}\right]}{2TR\left(1-e^{-X}\right)},\;\;
X=\frac{c\pi\beta\hbar}{R},\nonumber\\
& & U=\frac{c\pi\hbar N}{2R\left[e^{\frac{\beta c\pi\hbar}{2R}}-1\right]}+\frac{c\pi\hbar N}{R\left[e^{\frac{\beta c\pi\hbar}{R}}-1\right]}.
\label{6}\\
& & P=\frac{c\pi N\hbar\left[e^{-\frac{X}{2}}+3e^{-X}\right]}{2R^2\left(1-e^{-X}\right)}\frac{dR}{dV},
\label{7}
\end{eqnarray}
with $V$ the thermodynamic volume and  $dR/dV=3^{-1}{(3/(4\pi))}^{1/3} V^{-2/3}$. Note that in deriving the (\ref{6})
and (\ref{7}) (the (\ref{5}) is insensitive to the zero point energy), the zero point energy, as customary, has been sutrbacted
from $U$. It is believed that such a zero point energy becomes important at Planckian scales, where a possible non commutative quantum 
structure of the spacetime can emerge \cite{a3,a4,a5,a6}.  
From the (\ref{6}) and (\ref{7}) we see that $PV=U/3$, i.e. the spectrum formula (\ref{3}) depicts a radiation field made of
gravitons. Entropy and internal energy both depend on the temperature $T$, on the areal radius $R $ of 
the box and on the number of graviton's oscillators $N$. In a usual classical ideal gas the internal energy $U$ is a function only of the thermodynamic temperature $T$ and to obtain a sound thermodynamic limit we have $N\sim R^3\sim V$.
In our case, finite size effects are present and the internal energy is also a function of $R$. The spectrum 
(\ref{3}) becomes continuum only in the limit $R\rightarrow\infty$ and as a consequence in this limit quantum effects 
are negligible and $U=U(T)$. A physically interesting study is to consider physically reasonable expressions for the internal 
energy and then look to the resulting expression for $T$ by inverting the equation (\ref{6}) and for 
$S$. Since the well known
behavior for the BH temperature $T_{BH}$
\begin{equation}
T_{BH}=\frac{c\hbar}{4\pi K_B R},
\label{8}
\end{equation}
derives from $U\sim R$, it is interesting to see the modification of $T=T(R,N)$ due to a given reasonable choice for $U$.
In sections 3,4 we consider the 'classical case' with $U\sim R^k,\;k\geq 1$, while in section 5 we analyze the corrections arising
from quantum fluctuations.

\section{Temperature and entropy for trapped gravitons: classical viable cases}

To start with, we invert the expression (\ref{6}) with the only acceptable root given by
\begin{equation}
K_B T=\frac{c\pi\hbar}
{2R\ln\left[\frac{Q}{4}+\frac{1}{2}\sqrt{\frac{Q^2}{4}+6Q+4}\right]},\;\;\;\;Q=\frac{c\pi\hbar N}{RU}.
\label{9}
\end{equation}
Note that in the limit with $Q^2<<1$, we have $U=2NKT$, in contrast with usual ideal gas for which
we have $U=\frac{3}{2}NKT$.

Concerning the internal energy $U$, it is natural, to a classical level, 
to take $U=Mc^2$, where $M$ is the ADM mass of the graviton's 'star' and introduce, 
as customary, the density profile $\rho=\rho(r)$, where as usual
\begin{equation}
U(R)=4\pi c^2\int_0^R\rho(r)r^2 dr. 
\label{11}
\end{equation}
In practice, to massless gravitons with energy $E_{\ell n}=\hbar{\omega}_{\ell n}$ we associate a 'weight'
$m=E_{\ell n}/c^2$. Moreover, we are not interested in the study of the metric inside the spherical box but 
rather on the thermodynamic features of the radiation's ball of gravitons. From the astrophysical point of view,
this study could be of interest, for example, in the modeling of exotic but interesting objects as the supposed gravastars
\footnote{Gravastars are considered as an alternative to BH.}
\cite{17,18} where a thin shell made  with the cosmological constant can trap gravitons inside.\\
For a physically reasonable and simple expression for $\rho(r)$ we can choose
$\rho(r)=\frac{q}{r^k},\;\;q\geq 0$ and $k<3$ to guarantee the convergence of (\ref{11}). In particular, for
$\rho\sim 1/r$ we have $U\sim R^2$, for $\rho(r)=q$ we have $U\sim R^3$, while for $\rho\sim 1/r^2$ we have
$U\sim R$. Obviously we could consider exotic expressions with $k$ assuming non-integer values (fractal-like  
case), but the cited above expressions for $\rho$ are the most frequently used to model the interior of a star.
To assure a good thermodynamic limit, generally we must pose $N=sR^3,\;s\in\Re^+$. Only in the BH case, according to
the holographic nature of a BH, we have $N=sR^2$. Also note that physical reasonability (stability)
requires a non-increasing expression for $\rho$ and as a consequence we do not study the cases with $U\sim R^k,\;k>3$.

\subsection{Case with $U\sim R^3$}

Firstly, we consider the case with $U=qR^3$. This corresponds to the choice $\rho(R)=k,\;k\in \Re^+$. This equation of state is often used in astrophysics to model a star with a dense core as a neutron star. From the (\ref{9}) and after setting, as usual,
$N=sR^3,\;s\in \Re^+$, we obtain
\begin{equation}
K_B T=\frac{c\pi\hbar}
{2R\ln\left[\frac{c\pi\hbar s}{4qR}+\frac{1}{2}
\sqrt{\frac{c^2{\pi}^2{\hbar}^2 s^2}{4R^2 q^2}+\frac{6c\pi\hbar s}{R q}+4}\right]}.
\label{12}
\end{equation}
In the limit $R\rightarrow 0$ we have $T\rightarrow +\infty$, similarly to the BH physics with $S\rightarrow 0$.
In this limit the entropy looks like:
\begin{equation}
S={\alpha}_1 A^2 + {\alpha}_2 A^2\ln(1/A^2)++o(1),\;\;\{{\alpha}_1,{\alpha}_2\}\in \Re^+.
\label{13}
\end{equation}
In the opposite limit, $R\rightarrow +\infty$, we have $K_B T\rightarrow q/(2s)$.\\
Remember that  for a BH in the same limit we have $T\rightarrow 0$. This first example shows some analogy but also  differences with the BH case. In particular, by supposing that $N=s R^2$ as in the BH case, the entropy also for small values of
$R$ where quantum effects are strong, looks like $S\sim V$. Also notice that the temperature (\ref{12}) presents the behavior
$\sim 1/R$ suitable for a BH but with the term given by a logarithm at the denominator of the (\ref{12}).

\subsection{Case with $U\sim R^2$}

The case $U=q R^2$ can be obtained with ${\rho\sim 1/r}$. This is a diverging density but giving a finite expression for $U$.
We expect that in this case formulas are more closed related to the BH ones since we are approaching the behavior 
$U\sim R$ suitable for a BH. With $U=qR^2$ and $N=sR^3$ we have
\begin{equation}
K_B T=\frac{c\pi\hbar}
{2R\ln\left[\frac{c\pi\hbar s}{q}+\frac{1}{2}
	\sqrt{\frac{c^2{\pi}^2{\hbar}^2 s^2}{4q^2}+\frac{6c\pi\hbar s}{q}+4}\right]}.
\label{14}
\end{equation} 
Since $q,s$ are positive constant, we obtain the interesting result that in this case the temperature is proportional to 
$T_{BH}$, given by (\ref{8}), by a constant term. Moreover, we have $S\sim V$, i.e. the entropy scale as the thermodynamic volume.
This means that a ball filled with gravitons with $U\sim R^2$ approaches the BH behavior, but with the fundamental 
difference that the 
entropy scales as the volume rather than the area.
We can choose the parameters $s,q$ in such a way that $T=T_{BH}$. This can be obtained by setting
\begin{equation}
\frac{s}{q}=\frac{2\left(e^{4{\pi}^2}-1\right)}{c\pi\hbar\left(e^{2{\pi}^2}+3\right)}.
\label{15}
\end{equation}
Hence, an hypothetical extreme astrophysical object made of trapped gravitons with the (\ref{15}) would emit a (cold) thermal radiation at 
the same temperature of the event horizon of a BH. However note that it is the relation between the entropy and the area of the event horizon that is the heart of the BH physics showing the holographic nature of the BH 's gravity.
The result is that, by assuming the usual behavior $N=sR^3$, we approach the proportionality $S\sim A$ by approaching the 
behavior $U\sim R$. 

\subsection{Case with $U\sim R$}
In this subsection we analyze the BH-like behavior $U=qR$. This behavior can be obtained from the (\ref{11}) with
$\rho(r)\sim 1/r^2$. This density profile is typically used to model the huge core of neutron stars.
By supposing that the configuration of the graviton's radiation ball it is not in a BH state, after posing
$\rho(r)=q/(4\pi r^2)$ and remembering that in the BH case we have $U=c^4 R/(2G)$, we must have
\begin{equation}
qR=Mc^2,\;\;\;\;R>\frac{2GM}{c^2}\rightarrow q<\frac{c^4}{2G}.
\label{16}
\end{equation}
First of all, note that in the limit $R\rightarrow 0$ we have $T\rightarrow +\infty$ and
$S\rightarrow 0$, as happens for a BH.
In the opposite limit, $R\rightarrow +\infty$ we have $T\rightarrow 0$. Also in 
the limit $R\rightarrow\infty$ we have,
always with $N=s R^3$,
\begin{equation}
K_B T=\frac{c\pi\hbar}{2R\ln\left(\frac{c\pi\hbar sR}{2q}\right)}.
\label{17}
\end{equation} 
For $S$ we have the behavior
\begin{equation}
\frac{S}{K_B}=\frac{q}{2c{\pi^2\hbar}}A+\frac{qA}{2c\pi\hbar}
	\ln\left(\frac{sc\sqrt{\pi}\hbar\sqrt{A}}{4q}\right)+corrections.
\label{18}
\end{equation}
In the leading term of (\ref{18}) does appear the term proportional to $A\ln(A)$ with 
the usual BH behavior appearing as a correction to the leading term.
Note that this behavior also happens by taking $q=c^4/(2G)$ instead of the
inequality (\ref{16}). This shows that the fundamental ingredient to obtain the BH formula is provided by the 
holographic relation $N=sR^2$. These reasonings suggest that a BH genesis can be seen as a physical process where matter becomes more and more dense. When inequality (\ref{16}) is saturated, a BH emerges only when matter-energy inside the event horizon
condensates near the event horizon.
Since no usual matter-energy can live inside a BH (at least static), we may suppose that matter eventually falling inside the event horizon
behaves like a radiation field.

\section{The black hole case}

A remarkable debated question regards the origin of the degrees of freedom leading to the BH entropy. In this paper we use the 
view that the degrees of freedom are present inside the horizon as massless gravitons forming a discrete spectrum
(as far as $R$ is not large 'enough' to justify the continuum approximation). This could be a practical
realization of the 
{\it BH complementarity conjecture} advocated by 't Hooft and Susskind (see \cite{Page} and references therein), where two copies of
information are stored in the Hawking radiation and inside a BH (gravitons in our case). 

In the examples of section 3, we obtained physically reasonable models with $N=sR^3$. Note that for any choice of the
internal energy $U=U(R)$, we can always choice an 'exotic' distribution $N=RU(R)$ in order to obtain for $T$ the behavior given by
(\ref{8}). In any case, there exists only a possibility to obtain the behavior (\ref{8}) with $S\sim A$, i.e. by setting
the internal energy given by the ADM mass-energy $U=c^4R/(2G)$ with the holographic prescription $N=sR^2$.
With these choices we have \cite{1}

\begin{eqnarray}
& & R=\sqrt{\pi N}L_P\sqrt{\frac{1}{\left[e^{\frac{{2\pi}^2}{\alpha}}-1\right]}+
	\frac{2}{\left[e^{\frac{4{\pi}^2}{\alpha}}-1\right]}},\\
\label{19}
& & S=K_B Y(\alpha)\frac{A_h}{4L_P^2},\;\;A_h=4\pi R^2,\label{20}\\
& & Y(\alpha)=\frac{b}{\alpha\pi^2\left(3+e^{\frac{2\pi^2}{\alpha}}\right)}\nonumber\\
& & b=-\alpha e^{\frac{4\pi^2}{\alpha}}\ln\left(1-e^{-\frac{2\pi^2}{\alpha}}\right)-
\alpha e^{\frac{4\pi^2}{\alpha}}\ln\left(1-e^{-\frac{4\pi^2}{\alpha}}\right)+6\pi^2+\nonumber\\
& & +2\pi^2 e^{\frac{2\pi^2}{\alpha}}+\alpha \ln\left(1-e^{-\frac{2\pi^2}{\alpha}}\right)+
\alpha\ln\left(1-e^{-\frac{4\pi^2}{\alpha}}\right). 
\end{eqnarray}
In \cite{1} we have shown that, in order to obtain the BH entropy, a temperature looking like
\begin{equation}
T=\alpha T_{BH}=\frac{\alpha c\hbar}{4\pi K_B R},\;\;\alpha\in (0,\infty),
\label{21}
\end{equation}	
with the only solution $\alpha\simeq 2.2$ leads to the BH formula $S=S_{BH}=\frac{K_B A}{4L_P^2}$.
This result is far from trivial. 
In fact, in the BH case we impose a constraint on the entropy, i.e. $S=S_{BH}$. Since of the distribution (\ref{4}), entropy in
(\ref{20}) is a complicated function of $\alpha$. As a consequence, the equation $Y(\alpha)=1$ is not trivial. 
In particular, this equation is sensitive to the choice made for ${\omega}_{\ell n}$ given by the (\ref{3}). With a different choice,
a different value for $\alpha$ can be generally found. As an example, in \cite{1} we have considered another model with 
${\omega}_{\ell n}$ given by the (\ref{3}) but with the partition function summed up with respect to the azimuthal index
$m$. As a result we obtained $\alpha\simeq 8.48$. More generally we have an equation with $Y$ depending on the distribution
chosen for ${\omega}_{\ell n}$, i.e. $Y=Y_{\{{\omega}_{\ell n}\}}(\alpha)=1$. The reader can point out
that the important fact is that 
a solution for the equation $Y=1$ exists. Fortunately, this is not the case. In fact, the spectrum (\ref{3})
leads to a radiation field with $PV=U/3$.
If we consider the entropy, $S_{BH}=\frac{K_B A}{4L_P^2}$, with $T$ given by the (\ref{21}), from the first law 
$TdS_{BH}=dU+PdV$ integrated with respect to $R$ we obtain:
\begin{equation}
PV=(\alpha-1)\frac{U}{3}.
\label{22}
\end{equation}
First of all, note that with the value $\alpha=1$, i.e. $T=T_{BH}$, we have the relations $P=0$ with $T_{BH}dS_{BH}=dU$
suitable for the exterior of
a  Schwarzschild BH. This implies that a static BH is pressureless outside the event horizon following the usual
treatment. Conversely, within our treatment, a pressure
radiation term naturally arises.\\   
The formula (\ref{22}) does imply that to obtain a radiation field for the interior of a BH we must impose
exactly $\alpha=2$.  Since the value we obtained is closed to the expected value $\alpha=2$, we can reasonably consider our 
approximate formula (\ref{3}) as a good approximation, also showing that, thanks to the unacceptable solution
$\alpha\simeq 8.48$, our choice present in \cite{1} to promote only $n$ to a quantum number looks in the right direction.\\
Note that the formula (\ref{3}) gives the zeros of the Bessel functions $j_{\ell}(kr)$ with an error quickly decreasing.
For example, the first zero is correct within the $4\%$, while the second zero is exactly within a percentual less than 
$1\%$. Hence the obtained error $(2.2-2)/2\sim 10\%$ is expected within the approximations made.\\
Summarizing, the equation $Y=1$ imposed to obtain $S=S_{BH}$ strongly depends on the spectrum chosen for ${\omega}_{\ell n}$
and our approximate formula (\ref{3}) is a good starting point to explore the thermodynamic of trapped gravitons.
Any physically acceptable approximation for $\{\omega\}$ should give a value for $\alpha\simeq 2$.

As a final consideration for this section,
note that formula (\ref{22}) with $\alpha=-2$ it gives a dark energy distribution with the usual equation of state $PV=-U$. The fact that a negative value for $\alpha$ but with the same $|\alpha|=2$ it gives a dark energy configuration is an intriguing fact.
However, in this case we have $T<0$ and the statistical study of this paper and in\cite{1} are obviously not suitable.
We may think to take $T<0$ and $U<0$ inside the black hole, but we get the solution $\alpha=-2$ but again with radiation
$PV=U/3$ and with $P<0$ and $T<0$. These considerations can be of interest in line of the papers \cite{V2,V3} where has been 
hypothesized that the cosmological constant can be made of gravitons near a BEC state. 
This can certainly be matter for further investigations, but the fact that equation (\ref{22}) admits the BH solution $S=S_{BH}$ with
$|\alpha|=2$ representing radiation ($\alpha=2$) and dark energy $\alpha=-2$ is an intriguing fact.

\section{Building corrections to the BH entropy}

As well known, the determination of the 
corrections to the semi-classical BH entropy is an important ingredient that should be incorporated in any
consistent quantum gravity theory.\\
As well known \cite{c1}, for a given thermodynamically stable system (positive specific heat $C$) with entropy $S_0$,
thermal fluctuations generally lead to the following corrections $S_c$ for the entropy:
$S_c=S_0+1/2\ln(S)$. Thus logarithmic corrections hold for any stable thermodynamic systems. In practice,
the logarithmic corrections for a BH can be seen as due to finite size corrections for a large radius $R$.
Nevertheless, for a BH ($C<0$) quantum fluctuations  are present \cite{c1} and logarithmic corrections
are also expected to arise due to quantum fluctuations of the geometry. Thermal fluctuations generally produce a positive 
logarithmic correction to the BH entropy, while quantum ones generate negative logatithmic corrections \cite{f1}.
These corrections are in competition. For example, for BTZ black holes (see \cite{f1} and references therein), thermal and quantum logarithmic corrections cancel out. Note that asymptotically flat BH's are thermodynamically unstable. For this reason,  
a BH can be seen as thermodynamically immersed in a spherical isothermal cavity (see for example \cite{g4}
and references therein) and as a result 
for an asymptotically flat BH with entropy $S_{0BH}$ we thus have 
$S_{BH}=S_{0BH}-{\alpha}_0\ln(S_{0BH})$. The constant ${\alpha}_0$ is model dependent and ${\alpha}=3/2$ in a string context
\cite{11,11b,c2} and ${\alpha}_0=1/2$ in loop context \cite{ad4}.

It is thus generally believed that \cite{11,11b,g4,c2,ad2,ad3,ad4,g1,g2,g3,g5}) quantum and thermal 
fluctuations generate logarithmic corrections to the BH entropy formula with other corrections given by:
\begin{equation}
S_{BH}=\frac{A}{4L_P^2}-a_0\ln\left(\frac{A}{4L_p^2}\right)+higher\:\:order\;\;corrections\;.
\label{c1}
\end{equation}
The formula (\ref{c1}) is expected to hold for large $R$ and the positive constant depends on the different kinds of derivations.
Other kinds of corrections can be found for example in the context of entangled
entropy (see for example \cite{C1} for a recent review and references therein) such as
\begin{equation}
S_{BH}=\frac{A}{4L_P^2}+a_1\sqrt{A}+a_2+a_3\frac{1}{\sqrt{A}}+o(1/\sqrt{A}),\;\;\;\{a_1,a_2,a_3\}\in\Re,
\label{c2}
\end{equation}
or a mixture between (\ref{c1}) and (\ref{c2}) motivated \cite{C2} by Planckian effects caused by the
generalized uncertainty principle (GUP) \cite{g1,g2}. In the limit of large areal radius  $R$, the semi-classical expression is regained.

As noticed in \cite{Page}, the calculations of the logarithmic corrections to the BH entropy are physically
ambiguous. In fact, as stated above,
thermal fluctuations always lead to logarithmic (positive) corrections and as a consequence
it is not easy to calculate the contribution caused by quantum fluctuations or ask if cancellation do occur.
Since the specific heat $C$ for a BH is negative, thermal fluctuations cannot be calculated 
in the usual canonical description around a given equilibrium configuration, since in this case the canonical energy acquires
an imaginary part \cite{f1} and only a microcanonical description is available with fixed radius $R$.
These reasonings show that for large asymptotically flat BHs, the determination of thermal fluctuations by the usual saddle point
technique at the equilibrium point is not suitable. Only at the UV scales, near the Planck scales
it is expected that quantum fluctuations induced by the quantum fluctuations of the geometry become dominant. 
This is the main reason for which we make
the physical distinction between IR (i.e. macroscopic) scales and UV (Planckian) scales. Planckian corrections can be calculated
by introducing a mathematical and physical description suitable for a non-commutative spacetime that is expected to emerge
approaching the Planck scales. For the modified temperature we can choose
\footnote{The modification \ref{c3} is often used in 
	literature also in models with a Planckian modified dispersion relation for spherical configurations  
	motivated by the generalized uncertainty principle (GUP).}
\begin{equation}
T_{BH}=\frac{c\alpha\hbar}{4\pi K_B R}+\frac{B}{R^2}+\frac{F}{R^3}+higher\;\;orders,\;\;\;\{B,F\}\in\Re^+.
\label{c3}
\end{equation}
In the context of the GUP \cite{g2,g3,g5}, it is often used an expansion similar to the (\ref{c3}) but with $B=0$.
In both cases, for guidance, no experimental data are at our disposal
and only physically reasonable although speculative conjectures can be invoked, although string and loop quantum 
gravity can provide a reasonable guidance. 

It is natural to ask if corrections to the semi-classical BH formula can be obtained within our approach. In this framework, 
conversely with respect to the strategy used in the sections above, we cannot generally derive an expression for 
$U(R)\sim 1/R$ in terms of the (\ref{11}) since of the requested convergence of the integral (\ref{11}).
To this purpose, we study separately the IR and UV Planckian scales.

\subsection{IR limit}

This is the limit of very large volumes. In this regime it is also expected that the logarithmic correction arises as a 
leading correction. As noticed above
\cite{Page}, it is not a simple task to separate contributions to the BH corrections arising from
thermal and quantum gravity effects.\\
Our approach is statistical and our starting point are formulas (\ref{4})-(\ref{6}). 
Moreover, as shown in the sections above, an expression for $U=Mc^2$ in the 'classical' case can be represented  in terms of a given 
effective graviton's density. By considering thermal corrections, 
we expect that $U=Mc^2+\delta M c^2$, where the term 
$\delta M$ indicates fluctuations. It is expected that the expression chosen for $\delta M$ as a function
of the size $R$ of the BH's horizon  
depends on the model chosen to calculate such fluctuations. In this regard, our approach is 
phenomenological and the form of the fluctuations is model dependent. Hence, 
according with our procedure, we can assume the entropy formula (\ref{5}) with the statistical expression for 
$U$ given by the (\ref{6}) and therefore choose the modified expression for $U(R)$ leading to the corrections to the leading order.
In this regard, note that we can always choose the internal energy $U$ and the Hawking temperature in the formulas
(\ref{5}) and (\ref{6}) in such a way that both logarithmic and 'entangled' corrections do arise.
Concerning the corrections $\sim {\sqrt{A}}$, these can be easily obtained in the large volume limit by adding corrections 
$U$ looking like $\sim 1/R$ with $U=\frac{c^4 R}{2G}+\frac{U_c}{R},\;U_c\in\Re$
together with the modified BH temperature (\ref{8}) 
looking like the (\ref{c3}).
Corrections $\sim\sqrt{A}$ follow by a series expansion in terms of $1/R$ in (\ref{5}).\\
Logarithmic corrections to the BH entropy in the IR limit, i.e. looking like $\sim\ln(A)$, 
can be obtained by taking
\begin{equation}
U=\frac{c^4 R}{2G}+\frac{{\alpha}_1}{R}\ln\left(R\right)+\;higher\;\;\;order\;\;\;corrections,
\;\;\;\;{\alpha}_1\in\Re.
\label{cc3}
\end{equation}
A mixing between logarithmic corrections and the ones looking like $\sim{(\sqrt{A})^z}$ (with $z$ an integer $\leq 1$) can be obtained 
by adding to the (\ref{cc3}) terms proportional $\sim 1/R^n,\;n\in\mathbb{N}-\{0\}$ and using the 
(\ref{c3}).\\
The 'phenomenological' correction $U_c\sim \ln(R)/R$ in (\ref{cc3}) can be justified in the following way.
The microcanonical entropy is 
given by $S_{BH}=K_B\ln\Gamma(E)$, where as usual $\Gamma$ denotes the counting of the allowed states.
Since for a system the function $\Gamma$ is negligible out of the equilibrium\footnote{In our case one may think that
	graviton's radiation inside the BH is in equilibrium with the event horizon that acts as the cavity of a 
	black body radiation.} configuration $E^{\star}$,
we have, from statistical mechanics textbooks, that fluctuations around $E^{\star}$ for $\ln\Gamma$ look like
$\ln(E)$ and since for a BH $U\sim ST$ and $E\sim R$, this motivates the correction for $U$ in the (\ref{cc3}).
Obviously the limit of our approach is that we cannot calculate the exact expression for ${\alpha}_i$ in
(\ref{cc3}). However, our 'phenomenological' approach can give a statistical description once a physically
motivated expression for $U$ and $T$ are chosen, motivated for example by a quantum gravity model. 

\subsection{UV Planckian limit}

The situation is rather more intriguing in the UV limit. In particular, corrections (\ref{c2}) and (\ref{c3}) are often obtained 
in the literature by  
reasonings regarding quantum gravity effects although 
a full understood shared quantum gravity theory is still lacking and
the form of the subleading corrections at small scales is debated. 
It is thus 
interesting to ask what are possible physically motivated expressions of the leading term in the UV Planckian limit. 
These expansions in powers of $m_{pl}/M$ (where $m_{pl}$ is the
Planck mass), 
in order to obtain corrections to the semi-classical BH entropy, 
are used in the GUP context\cite{ad1,ad2,ad3,ad4,g2,g3,g5}. The corrections so obtained are thus certainly suitable 
at IR scales with $R$ of the dimension of a macroscopic object 
but the convergence of the series expansion should be checked at Planckian scales where $R\sim L_P$.\\ 
First of all, note that after including the zero point term 
(subtracted in the (\ref{6})) given by $U_0=2c\pi\hbar N/R$ in the (\ref{6}), we have that the equation 
$Y(\alpha)=1$ has not solutions. In particular, the maximum value for the entropy is obtained with $\alpha\simeq 40$
with $S\simeq 0.04 S_{BH}$.  
This can be related to the discussion of section 4 concerning the dependence of the equation $Y(\alpha)=1$ on the distribution 
chosen. In the following, as usual, we continue to subtract the zero point energy from the (\ref{6}).\\
To start with, we must choose physically motivated modifications for $T_{BH}$ and $U$. In the context of the so called GUP, the 
following heuristic modification (deformation) of the usual position-momentum $x-p$ uncertainties relations\footnote{See for example
also \cite{g2} and references therein for possible modifications.}
, motivated by the BH physics, has been suggested
\cite{g1}:
\begin{equation}
\Delta x\Delta p\geq \frac{\hbar}{2}\left[1+{\gamma}^2{\Delta}^2 p\right],
\label{uv1} 
\end{equation}
where the symbol $\Delta A$ denotes the uncertainty of some operator $A$ in a quantum state $\omega$. Formula (\ref{uv1})
implies a minimal uncertainty for the position operator of the order of the Planck length, provided that the constant
$\gamma$ is chosen in a suitable way. Since for a BH we have $c^2\Delta M\sim c\Delta p$ and since it is expected that for a BH
the (\ref{uv1}) is satured with $\Delta x\sim\Delta R$, this motivates the corrections to $U$ looking like $1/R$. Note that many versions of GUP exist in the literature, see for example \cite{15} for a recent new proposal.
The drawback of the formula (\ref{uv1}) and of similar proposals present in literature,
is that the (\ref{uv1}) implies a 'deformation' of the usual dispersion relations for a relativistic
particle. At the present time, no such a modification has been observed in the real word. Moreover, notice that in the GUP (\ref{uv1})
it is implicitly assumed (see \cite{a8} for a nice review and also \cite{a5}) that 
$\Delta x\sim\Delta y\sim\Delta z\sim\Delta R$ for
the cartesian coordinates $\{x,y,z\}$,
i.e. that a spherical symmetry is implicitly assumed. In \cite{a3} it has been shown that a non-commutative spacetime at the Planck length preserving Lorentz covariance can be obtained starting from physically motivated uncertainty relations (generalized in \cite{a4} using Penrose's inequality). These uncertainty relations, in the spherical case, reduce to \cite{a4} 
\begin{equation}
\Delta R\geq s L_P,\;\;\;c\Delta t\Delta R\geq s^2L_P^2,\;\;s\sim 1.
\label{uv2}
\end{equation}
Hence, if spherical symmetry is assumed with $c\Delta t\sim \Delta R$, a minimal uncertainty for $\Delta R$ arises for a non-commutative spacetime, according with the GUP expression (\ref{uv1}). A remarkable feature of the model present in 
(\cite{a3}) is that also a minimal area \cite{a7} (see also \cite{a4}) and volume do appear, also for localizing states
$\omega$ that have not spherical symmetry. As a consequence of these reasonings motived by the DFR non-commutative model present in
\cite{a3}, the following expansion for $U$ can be written:
\begin{equation}
U=\frac{c^4}{2G}R+\frac{C_1 c^4 L_P^2}{2GR}+\frac{C_2 c^4L_P^3}{2GR^2}+\cdots
+\frac{C_n c^4 L_P^{n+1}}{2GR^n},\;\;\;\;\{C_n\}\in\Re^+,
\label{c6}
\end{equation}
The (\ref{c6}) can be also justified by the GUP (\ref{uv1}). In fact, we can solve the (\ref{uv1}) for 
$\Delta p$ with $\Delta p\sim 1/\Delta x+{\Delta p}^2/\Delta x,\;\Delta x\sim \Delta R$ and iteratively obtain a series expansion
in terms of powers of $1/R$ and this motivate the (\ref{c6}). However, as explained in
this section, GUP has been obtained with the implicit assumption of spherical symmetry and in this regard it is equivalent to the
physically motivated spacetime uncertainty relations (\ref{uv2}) for a non-commutative spacetime.
Note that 
we have $R_{min}\sim s L_P$, and all terms look like $\sim L_P$ at the Planck length with the magnitude depending on the 
dimensionless constant $\{C_n\}$. A similar expansion can be written for $T_{BH}$ as in (\ref{c3}).\\
For a pedagogical application, consider the expansion (\ref{c6}) truncated to the term $\sim 1/R^2$ together with the 
(\ref{c3}) up to the term $\sim 1/R^2$. For the corrected BH entropy at the leading order by formally expanding in the 
extreme UV limit
$R\rightarrow 0$ we have:
\begin{equation}
S_{BH}=-\frac{c\hbar C_2 L_P}{4B}\ln\left(\frac{c^2{\pi}^2{\hbar}^2 R^2}{2 K_B^2 B^2}\right)+
\frac{c\hbar C_2 L_P}{2B}+o(1),
\label{c5}
\end{equation} 
where obviously in our model the constant $C_2,Q$ are left arbitrary. We thus obtain a $\ln(A)$ as a leading correction
in the extreme UV limit for $R\rightarrow 0$. In the usual approach with GUP \cite{g6, g7} and according with results 
obtained in string and loop contexts, only odd powers in $1/R$ are present in (\ref{c3}) and (\ref{c6}). In any case, at any 
chosen 'odd' truncation  $1/R^{2n+1}$ for (\ref{c3}) and (\ref{c6}), the (\ref{c5}) still holds.
By considering, for example, the terms $\sim 1/R^3$ in (\ref{c3}), we obtain an expression very similar to 
(\ref{c5}) but with $C_3$ instead of $C_2$ and $C$ instead of $B$. 
From our simple setups, we also found $\ln(A)$ corrections to the BH entropy formula but with a 
Planckian nature and extrapolated for $R\rightarrow 0$. The leading logarithmic 
term in the (\ref{c5}), according to the (\ref{uv2}), give a positive contribution for:
\begin{equation}
B\geq \frac{c\pi\hbar s L_P}{\sqrt{2}K_B}=\overline{B}.
\label{ce}
\end{equation}
By setting $B\geq \overline{B}/e$, we obtain $S\geq 0$ in the limit $R\rightarrow 0$.
It is important to stress again that the
(\ref{c5}) has been obtained in the extreme (pedagogical)
UV limit $R\rightarrow 0$. In this pedagogical limit (without
invoking non-commutative Planckian effects) obviously the singularity is reached and as a result in this limit
we have $\{T_{BH}, U,S_{BH}\}\rightarrow +\infty$ \footnote{This can be seen as a kind of BH version of the
	famous 'ultraviolet catastrophe' in the black body radiation context.}.  
If we restrict to the behaviors at 
$s L_P\leq R<\infty$, as requested by a quantum spacetime motivated model, we can obtain results in agreement with the
(\ref{c1}) by choosing appropriate modified expressions for 
$U$ and $T$ motivated by non-commutative effects and thermal fluctuations. 
In particular, we can invert the problem: take the expression (\ref{c1}) as starting point, and looks to the modified expressions
for $T$ and $U$.
This can be matter for the further investigation.

Summarizing, at IR macroscopic scales logarithmic corrections can be allowed in our model, but, as pointed in \cite{Page}, their 
quantum or thermal nature is questionable. 
Quantum corrections arising from quantum fluctuations of the geometry
are expected to be dominant at Planckian scales. Within our model and procedure, by considering 
modified expression for $U$ and $T_{BH}$ motivated by non-commutative effects at such scales, the usual $\ln(A)$ expression
arises as a correction in the range $s L_P\leq R<\infty$ (See the conclusions for more details). 
The important fact is that our approach, with gravitons inside the BH providing the 
degrees of freedom leading to the BH entropy, according to the
{\it Bohr BH complementarity conjecture} advocated by 't Hooft and Susskind, permits one to obtain corrections to the BH entropy 
from a statistical point of view, with the underlying quantum gravity theory modeling the modified expressions for
$U$ and $T_{BH}$. Our model predicts, thanks to the (\ref{6}), the behavior of the
function $N=N(R)$. Also in this case, we can obtain the UV pedagogical
limit for $R\rightarrow 0$, i.e. $N(R\rightarrow 0)$. To this purpose, in the UV limit we can write the (\ref{c3}) in a form with the explicit
dependence on the Planck length $L_P$:
\begin{equation}
T_{BH}=\frac{c\alpha\hbar}{4\pi K_B R}+\frac{c\hbar K_1 L_P}{4\pi K_B R^2}+\frac{c\hbar K_2 L_P^2}{4\pi K_B R^3}+
\cdots+\frac{c\hbar K_n L_P^n}{4\pi K_B R^{n+1}},\;\;\;\{K_n\}\in\Re^+.
\label{c9}
\end{equation}
After taking the first correction terms in the (\ref{c9}) and (\ref{c6}) we obtain, in the extreme UV limit 
$R\rightarrow 0$ \footnote{Note that to obtain a behavior $N\sim R^2+o(L_P/R)$, a Taylor expansion in terms of
	$L_P/R<<1$ must be performed, and this does apply also for the (\ref{c5}) in order to have
	$S_{BH}\sim A/4 +o(L_P/R)$}:
\begin{equation}
N \sim \frac{4\pi C_1 R}{K_1 L_P}+\frac{4\pi R^3}{K_1 L_P^3}+o(R^3).
\label{c10}
\end{equation}
Equation (\ref{c10}) clearly shows that the semi-classical behavior $N\sim R^2$, consequence of the holographic principle, does not hold
in the Planckian limit, i.e. semi-classical gravity is holographic but quantum fluctuations contrast holography.
Once again, remember that non commutative effects, thanks to the (\ref{uv2}), take operationally meaningless the measure of
lengths less than the Planck one and Taylor expansion must be stopped at $R/L_P\simeq s$. To this purpose,
singular terms $\sim 1/R$ obviously do appear in the (\ref{c10}) by taking higher-order corrections for $T_{BH}$ in the
(\ref{c9}), but this is not physically relevant since the UV 'catastrophe' is solved by quantum spacetime fluctuations.
The physically relevant fact is that in any case holography is destroyed by quantum geometry's fluctuations.

\section{A transition phase and BH evaporation}

Another interesting question concerns the study of the BH evaporation in our background.

In the usual derivation of the Hawking radiation \cite{2}, a strictly thermal radiation arises thanks to the
creation of pair's particles just outside the event horizon. Particles
with negative energy are captured inside the event horizon, while the ones with positive energy escape up to 
spatial infinity. As a result, a thermal radiation flux of particles will be measured by an observer at spatial infinity. This 
physical mechanism is practically left unchanged within our approach, but with the 
evaporation, as will be shown in this section, ending at some non-vanishing radius $R_c$.
In practice, the main difference with respect to the usual semi-classical calculations \cite{2} is quantitative and 
due to the corrections arising in the entropy expression, that in turn are caused by the modified internal energy 
and temperature given by 
(\ref{c6}) and (\ref{c9}) .\\
Hence, independently on the (important)
fact that our spectrum is discrete\footnote{To this purpose a continuum spectrum can be regained in the semi-classical  
	limit and thus it is expected that continuous approximation break down approaching the
	Planck scale.}, physically
we may think that when a quantum of negative energy $E=-\hbar|\omega|$, created near the event horizon, 
falls inside the event horizon, a graviton 
with energy $\hbar|\omega|$ annihilates with the incoming quanta with a decreases of the ADM BH's mass by a quantity
$E$.  
The intriguing fact of our model is that our model provides a physically reasonable (although obviously speculative)
mechanism to explain the fate of the degrees of freedom leading to the BH entropy, according to the {\it Bohr BH complementarity conjecture}, and also a reasonable mechanism to explain the fate of the negative energy particles falling inside a BH representing the incoming Hawking radiation.

To be more quantitative, we analyze the effects of our modified expressions for $U,T_{BH}, S_{BH}$ on the usual treatment 
of the evaporation process.\\ 
To start with, we study the
phenomenon with the modified entropy and the corrected internal energy $U$ given by (\ref{c6}) from a thermodynamics
perspective.\\  
As well known, the specific heat for a static BH is defined by
$\frac{dU}{dT}=-\frac{8\pi K_B G M^2}{c\hbar}$ and is strictly negative. This is a peculiarity of complex systems with long range
interactions. This does happen also by using the IR expression (\ref{cc3}), with terms looking like $1/R\ln(R)$ giving a
small correction leaving the expression for the specific heat $C=dU/dT$ negative. 
This situation changes drastically at Planckian regime, where in the expansion (\ref{c6}) terms looking like
$1/R$ start to become relevant. In this context, since from the (\ref{c3})
$dT/dR<0$, we have $C=dU/dT>0$, i.e. the specific heat becomes positive. This means that at Planckian scales the non-commutative
effects induced by the expected fuzzy geometry act as a repulsive force contrasting the attractive gravitational force.
This implies that there exists a critical radius $R_c$ such that $dU/dT=C=0$. Exactly at such a 
critical radius, a transition phase emerges denoting the transition between the IR and the UV Planckian scale.
From the (\ref{c6}), by assuming scales where the term $1/R$ is dominant, we found
\begin{equation}
\frac{dU}{dT}\sim \frac{c^4}{2G}\left(1-\frac{C_1}{R^2}L_P^2\right)\frac{dR}{dT}
\rightarrow\;\;R_c\simeq \sqrt{C_1} L_P.
\label{c7}
\end{equation}
On general grounds, we expect $C_1$ of the order of unity or greater and as a consequence
we expect the critical radius to be greater but near the Planck scale.
As a result, when the critical radius $R_c$ is reached during the evaporation process, the BH has
$C=0$ and becomes thermodynamically 'dead'. It is thus expected that evaporation stops when this critical radius is reached, by forming
the so called BH remnants. This picture is in agreement with the one GUP-motivated 
present in \cite{g2,g3,g4,g5,g6,g7} where a critical radius exists of the order of $L_P$ where $C=0$ and the BH becomes as a remnant
inert object. This fate could be related to the dark matter issue (see for example \cite{a4}). Moreover, in our model the specific
heat becomes positive near the limit $R\sim sL_P$, and this implies that the remnants configuration can be stable. In the usual 
approach with GUP, the configuration with $C=0$ is obtained exactly at the minimum radius, and the stability of the remnants 
so obtained is questionable.

The treatment above can be implemented by calculating the behavior of $R$ as a function of the time $t$. In this regard, we
adopt the technique in \cite{page}. In \cite{page}, the power $P$ of the Hawking radiation has been
computed by the
Stefan-Boltzmann law with $P=4\pi R^2\sigma T^4$ and 
$\sigma=\pi^2K_B^4/(60 \hbar^3 c^2)$. As a consequence, from the formula $P=-\frac{dU}{dt}$ and $U=Mc^2$ we have
\begin{equation}
\frac{dR}{dt}=-\frac{G\hbar^2}{\pi c^2 1920 R^2}.
\label{ev1}
\end{equation}
By inspection of the (\ref{ev1}), note that $dR/dt < 0$ and is diverging approaching $R=0$. Hence, equation (\ref{ev1})
can be integrated and also inverted up to $R=0$. After setting, for simplicity, the initial time $t_i=0$, 
at  $t=t_f$ with $R(t_i)=R_0$, we get:
\begin{equation}
t_f= K\left(R_0^3-R_f^3\right),\;\;\;K=\frac{640 \pi c^2}{G\hbar^2}.
\label{ev2}
\end{equation}
The finite evaporation time $t_{ev}$ can be calculated by setting $R_f=0$ in (\ref{ev2}). The semi-classical situation depicted by the
(\ref{ev2}) drastically changes in presence of the corrections obtained in this paper. For our purposes, it is sufficient
to consider expression (\ref{c6}), retaining the first term correction $\sim 1/R$ together the with the usual expression 
(\ref{8}) for the BH temperature $T_{BH}$. We obtain:
\begin{equation}
\frac{dR}{dt}=\frac{G\hbar^2}{\pi c^2 1920\left(C_1 L_P^2-R^2\right)}.
\label{ev3}
\end{equation}
From the (\ref{ev3}), note that the areal radius $R(t)$ is a decreasing function of $t$ for $R\in(\sqrt{C_1} L_P, +\infty)$,
is a non differentiable function at $R_c=\sqrt{C_1} L_P$, and increasing for $R<R_c$. This mathematical fact does imply that the differential equation (\ref{ev3}) can be integrated up to $R=R_c$, where the evaporation process ends
\footnote{In practice, the Cauchy problem with the initial condition at $R=R_0>R_c$ is well posed only up to $R=R_c$.}. We have:
\begin{equation}
t_f= 3K\left[\frac{1}{3}\left(R_0^3-R_f^3)-C_1 L_P^2\left(R_0-R_f\right)\right)\right],
\label{ev4}
\end{equation}
and $t(R_f)$ is defined and invertible in $R_f\in(R_c, R_0)$. The result (\ref{ev4}) is perfectly in agreement with the one obtained 
with (\ref{c7}). The vertical asymptote present in (\ref{ev3}) represents the mathematical translation of the fact that at
$R=R_c$ the BH is thermodynamically dead and cannot more change their dimensions. Note that, the inclusion of further terms
for $U$ and $T_{BH}$, given by the (\ref{c6}) and (\ref{c9}), does not change qualitatively the conclusions and their effect is to give 
corrections depending on $L_P^n$ to the (\ref{ev4}).

According to the results of section above concerning
a quantum spacetime, we can reasonable associate the correction terms in the (\ref{c6}) as due to the effective repulsive force
exerted by quantum fluctuations at the Planck scale.

These reasonings can be of interest also for the well known issue related to the loss information paradox due to the 
Hawking evaporation process. First of all, we may observe that the spectrum (\ref{3}) can be well approximated by a continuous
one for a large range of the radius $R$, and thus thermal nature of the Hawking radiation emerges. 
When approaching the Planck scale, we expect that the continuous approximation of (\ref{3})
breaks down and a non-thermal flux arises. However, the continuous limit remains an approximation, and with the discrete spectrum
(\ref{3}) the radiation flux is quasi-thermal and in some sense information could be regained. These reasonings are in line with the 
ones obtained in \cite{12,13,14,15}.
Moreover, the results given by (\ref{ev4}), according to a line of research present in the literature
(see for example \cite{g2,g3,g4,g5,g6,g7}),
show that quantum fluctuations stop the evaporation process. These two facts, in absence
of a sound quantum gravity theory,  certainly alleviate the information loss paradox.

Summarizing, our 'phenomenological' approach can reproduce under physically reasonable setups, standard results obtained in the 
literature regarding
the BH entropy corrections by assuming a non-commutative quantum spacetime at Planckian scales and with the benefit
of a simple description in terms of the statistical mechanics of gravitons inside the event horizon. Our conjecture is a simple 
but reasonable response to the question regarding the fate of any kind of matter-radiation falling inside the BH.

\section{Conclusions and outlooks}

In this paper we have studied the consequences of the proposal present in \cite{1} concerning the statistical
derivation of the semi-classical BH entropy in terms of trapped gravitons with a discrete spectrum. In particular, we studied the 
shape of the temperature and entropy in terms of the internal energy $U$ and the number $N$ of trapped gravitons forming $N$
harmonic oscillators. To explore differences with usual thermodynamics we have used 
$N\sim V$ in such a way that the usual thermodynamic
limit holds. For the internal energy $U$ we adopted the ADM mass-energy $M$ of the graviton's ball  with $U=Mc^2$.
The behavior of $U$ in terms of the radial radius $R$ of the confining box can be obtained by choosing a 
suitable 'density' $\rho(r)$ by means of the formula (\ref{11}). We show that the behavior $S\sim A$ can be approached by
setting density classical profiles that are diverging at $r=0$. Such density profiles are used to model the core of, for example,
neutron stars. However, the only ansatz leading to the BH formulas for $T$ and $S$ is provided by the choice
$U\sim R$ and $N\sim A$, according to the holographic nature of a BH. 
In practice, according to our proposal,  
any matter-energy falling inside a BH looks like an hologram made of radiation.

An important fact concerning the physical viability of our approach is that, as stressed in section 4 formula (\ref{22}), 
any viable spectrum formula for the radiation made of trapped gravitons inside a BH must satisfy
the equality $T=2T_{BH}$. As a result, any viable physically motivated approximate spectrum formula for
${\omega}$ must be, within the approximations made, near the theoretical value $\alpha=2$: this is what happens for our 
approximate spectrum formula (\ref{3}). Moreover, the statistical expressions for $S$ and $U$ strongly depend on the spectrum choice
(\ref{3}). As a numerical example, the partition function, obtained starting with the (\ref{3}) but summing up (see \cite{1})
with respect to the azimuthal index $m$ leads to $\alpha\simeq 8.48$, in complete disagreement with the value
$\alpha=2$ and then the Legendre index $m$ cannot promoted to a 'quantum' index as $n$. This means that our proposal in
\cite{1} to consider $\ell$ as a 'species' number rather than a quantum number is a
reasonable possibility. Moreover, this choice can solve the so called species problem since only massless (quadrupolar sextupolar..) gravitons can exist near and inside the event
horizon.

In section 5 we have also studied possible modifications arising to the semi-classical BH entropy formula by modifying in a physically reasonable way the expressions for $U$ and $T_{BH}$ arising from
thermal and quantum gravity corrections at Planckian and trans Planckian scales. We obtained in a simple manner
the usual logarithmic correction found for the semi-classical BH entropy 
both in the IR scale, where both quantum and thermal fluctuations arise, and in the UV limit, 
where Planckian physics is expected to 
dominate. The quantum modifications used in this paper
for $U$ are in agreement, for example, with the results present in \cite{ad1,ad2,ad3,ad4} where a mixing between corrections
$\sim 1/R$ and $\ln(R)/R$ are present in the IR limit, but can be
better motivate in the context of a non-commutative geometry \cite{a3,a4,a5,a6} that is expected at Planckian scales.
As a consequence of these facts, commented in section 6, a 
transition phase between the IR and the UV scales emerges, near the Planck one, where the specific heat becomes positive, i.e.
the BH becomes thermodynamically stable.
This shows that non-commutative effects at Planckian scales can be seen acting as a repulsive force, i.e. like an
effective cosmological constant. This phenomenon can be also found in papers (see for example 
\cite{g2,g3,g4,g5,g6,g7}) focusing on BH entropy corrections based on the well known GUP.

Motivated by the reasonings of section 5 regarding the IR the UV (also extreme) limits, an expression for $U(R)$, suitable for
$R/s L_P=T(R)>>1$ can be presented. Together with the (\ref{c9}) for $T_{BH}$ we have:
\begin{equation}
U=\frac{c^4}{2G}R-
{w}\frac{c^4 L_P^2}{2GR}\ln\left(\frac{R}{sL_P}\right)+
\frac{C_2 c^4L_P^3}{2GR^2}+\cdots
+\frac{C_n c^4 L_P^{n+1}}{2GR^n},\;\;\;\;s>1,\;\{C_n\}\in\Re.
\label{C1}
\end{equation}
Note that the constant $w$ is expected to be positive, i.e. quantum fluctuations are expected to dominate at such scales.\\
Also note that the log term in the (\ref{C1}) can be obviously written as
\begin{equation}
{w}\frac{c^4 L_P^2}{2GR}\ln\left(\frac{R}{sL_P}\right)=
w\frac{c^4 L_P^2}{2GR}\ln\left(\frac{R}{L_P}\right)-
w\frac{c^4 L_P^2}{2GR}\ln\left(s\right),
\label{C2}
\end{equation}
with the term $\sim 1/R$ restored as in the (\ref{c6}) and as a result the reasonings of section 6 are still valid.
Also note that the log correction is vanishing both at IR asymptotic scales and approaching a minimal uncertainty
\footnote{As noticed in \cite{a9}, the DFR model \cite{a3} does not allow a minimal uncertainty in the position, as
	requested by a self-adjoint operator, and this 
	happens only in the spherical case, according with GUP motivated results. Nevertheless, 
	a minimal length arises after that  
	a suitable quantum non-commutative field theory is built starting from the spacetime uncertainty relations.}
with the minimal length $R_{min}\sim sL_P$. This shows that the well known log corrections become polynomials at
$R\sim sL_P$. As shown in section 5.2, a log term also does arise in the UV pedagogical limit $R\rightarrow 0$.
It is easy to see that, after expanding the (\ref{c9}) and (\ref{C1}) in powers of $T(R)$ a mixing between the expressions
(\ref{c1}) and (\ref{c2}) are obtained. Moreover, extra corrections looking like the term $\ln(R/(sL_P))$ multiplied by a 
series expansion in terms of inverse powers of $T(R)$ are obtained. It is thus interesting to ask if 
effectively these corrections effectively do occur or
can be eliminated by a suitable relation between the coefficients in the series expansion
(\ref{c9}) or (\ref{C1}). This is matter for the next investigation, but it is clear from the reasonings above that 
our model is capable to obtain all corrections present in the literature with an oculate physically motivated choice for
$U$ and $T_{BH}$. 

A possible further line of research can be to obtain corrections to the BH entropy by using the generalized Gibbs-Boltzmann
distribution present in \cite{z1,z2}.

We have obtained the spectrum formula (\ref{3}) for a Schwarzschild BH. It is thus natural to ask
if an extension can be done for a Kerr-Newman BH. This is not a trivial task since of the complexity to obtain an analogue expression
for the Zerilli and Regge-Wheeler equtions in presence of a charged rotating configuration with angular velocity $J=a M$
\footnote{$M$ is the ADM mass.}
and an electric charge $q$.
A possible simple procedure may be to 'guess', inspired by the Newman-Janis algorithm \cite{NJ} and the technique present
in \cite{VVV}, the formula (\ref{3}). To start with, consider the temperature $T_{kn}$ of a Kerr-Newman BH given by
\begin{equation}
K_B T_{kn}=	\frac{c\hbar\left(r_+ -\frac{GM}{c^2}\right)}{2\pi(r_{+}^2+a^2)},
\label{23}
\end{equation} 
where $r_+$ denotes the location of the outer event horizon given by
\begin{equation}
r_+=\frac{GM}{c^2}+\sqrt{\frac{G^2 M^2}{c^4}-a^2-q^2}.
\label{24}
\end{equation}
To a first look, we may think to take formula (\ref{3}) with $R\rightarrow \sqrt{r_+^2+a^2}$, but this does not work
and both the temperature (\ref{23}) and the entropy $S_{kn}=K_BA/(4L_P^2)$, with $A=4\pi(r_+^2+a^2)$, cannot be obtained in this
'naive' manner. However, note that the temperature in (\ref{23}) is vanishing in the extreme case 
($G^2 M^2/c^4=a^2+q^2$) at the finite 
radius $r_+=GM/c^2$. Hence it is physically reasonable to expect that at zero temperature for a finite system must correspond
a zero internal energy and zero angular frequency, i.e. ${\omega}_{\ell n}(r_+=GM/c^2)=U(r_+=GM/c^2)=0$. As a consequence of these
reasonings, we get the following expressions:
\begin{eqnarray}
& & {\omega}_{\ell n}\simeq\left(2+\ell+2n\right)
\frac{c\pi\left(r_+ -\frac{GM}{c^2}\right)}{(r_{+}^2+a^2)}, \label{25}\\
& & U=\frac{c^4}{G}\left(r_+ -\frac{GM}{c^2}\right).\label{26}
\end{eqnarray}
With the expressions (\ref{25}) and (\ref{26}) we can obtain, after following the same procedure present in \cite{1},
the BH formula with the temperature proportional to (\ref{23}). The expressions of the static uncharged
case are regained in the limit
$a=q=0$.


\begin{thebibliography}{0}
\bibitem{1}S. Viaggiu,  Physica A {\bf 473} (2017) 412.
\bibitem{2}S. Hawking, Commun. Math. Phys. {\bf 43} (1975) 199.
\bibitem{3}L.D. Bekenstein, Phys. Rev. D {\bf 23} (1981) 287.
\bibitem{4}L. Bombelli, R. Koul, J. Lee, R. Sorkin,  Phys. Rev. D {\bf 34} (1986) 373.
\bibitem{5}V.P. Frolov, I. Novikov, Phys. Rev. D {\bf 48} (1993) 4545.
\bibitem{6}M. Srednicki, Phys. Rev. Lett.{\bf 71} (1993) 666.
\bibitem{7}R. Wald, Phys. Rev. D {\bf 48} (1993) 3427.
\bibitem{8}A. Strominger, C. Vafa, Phys. Lett B {\bf 104} (1996) 379.
\bibitem{9}V.P. Frolov, D.V. Fursaev, A.I. Zelnikov, Nucl. Phys. B {\bf 486} (1997) 339.
\bibitem{10}C. Rovelli, Phys. Rev. Lett. {\bf 77} (1996) 3288.
\bibitem{11}S. Carlip, Class Quantum Grav. {\bf 16} (1999) 3327.
\bibitem{11b}S. Carlip, Class Quantum Grav. {\bf 17} (2000) 4175.
\bibitem{12}C. Corda, Eur. Phys. J. C {\bf 73} (2013) 2665.
\bibitem{13}C. Corda, Class. Quantum Grav. {\bf 32} (2015) 195007.
\bibitem{14}C. Corda, Adv. High En. Phys. (2015) 867601. 
\bibitem{15}C. Corda,  Ann. Phys. {\bf 353} (2015) 71.
\bibitem{Page}D.N. Page New J. Phys. {\bf 7} (2005) 203.
\bibitem{16}S. Chandrasekhar, V. Ferrari, Proc. R. Soc. Lond. A {\bf 443} (1993) 445.
\bibitem{a3}S. Doplicher, K. Fredenhagen, J.E. Roberts, Comm. Math. Phys. {\bf 172} (1995) 187.
\bibitem{a4}D. Bahns, S. Doplicher, M. Morsella, P. Piacitelli, 
Advances in Algebraic Quantum Field Theory,(2015) 289-330 Springer. 
\bibitem{a5}L .Tomassini, S. Viaggiu, Class. Quantum Grav. {\bf 28} (2011) 075001.
\bibitem{a6}L. Tomassini, S. Viaggiu, Class. Quantum Grav. {\bf 31} (2014) 185001.
\bibitem{17}M. Bianchi, D. Grosso, R. Ruffini, Astron. Astrophys. {\bf 231} (1990) 301.
\bibitem{18}S. Viaggiu, Int. J. Mod. Phys. D {\bf 18} (2009) 275.
\bibitem{V2}S. Viaggiu, Gen. Relativ. Gravit. {\bf 48}:100 (2016).
\bibitem{V3}S. Viaggiu, Class. Quantum Grav. {\bf 34} (2017) 035018.
\bibitem{c1}S. Das, P. Majumdar, R.K. Bhaduri, Class. Quantum Grav. {\bf 19} (2002) 2355.
\bibitem{f1}A. Chatterjee, P. Majumadar, Phys. Rev. Lett. {\bf 92} (2004) 141301.
\bibitem{g4}M.M. Akbar, S. Das, Class. Quantum Grav. {\bf 21} (2004) 1383.
\bibitem{c2}R. Dijkgraaf, H. Verlinde, E. Verlinde,Nucl. Phys. B{\bf 371} (1992) 269.
\bibitem{ad2}M. Cavagli\'a, C. Ungarelli, Phys. Rev. D {\bf 61} (2000) 064019. 
\bibitem{ad3}M. Cavagli\'a, A. Fabbri, Phys. Rev. D {\bf 65} (2002) 044012.
\bibitem{ad4}R.G.L. Aragao, C.A.S. Silva, Gen. Relativ. Gravit. (2016) 48:83.
\bibitem{g1}D. Amati, M. Ciafaloni, G. Veneziano, Phys. Lett B {\bf 216} (1989) 41.
\bibitem{g2}Kim W, Son E J, Yoon M , JHEP {\bf 01} (2008) 035.
\bibitem{g3}F. Lenin, E. Herrera, E.A. Mena-Barboza,J. Torres-Arenas, Entropy {\bf 18} (2016) 406.
\bibitem{g5}Z.W. Feng, H.L. Li, X.T. Zu, S.Z. Yang, Eur. Phys. J. C (2016) 76:212.
\bibitem{C1}A. Barrau, C. Xiangyu, N. Karim, P. Alejandro, Phys. Rev. D {\bf 92} (2015) 124046.
\bibitem{C2}P. Bargueno, E. C. Vagenas, Phys. Lett. B {\bf 742} (2015) 15
\bibitem{ad1}M. Buric, V. Radovanovic, Class. Quantum Grav. {\bf 17} (2000) 33.
\bibitem{a8}G. Piacitelli, SIGMA {\bf 6} (2010) 073.
\bibitem{a7}D. Bahns, S. Doplicher, K. Fredenhagen, G. Piacitelli, Comm. Math. Phys. {\bf 308} (2011) 567.
\bibitem{g6}K. Nozari, A. S. Sefiedgar,  Phys. Lett B {\bf 635} (2006) 156.
\bibitem{g7}K. Nuicer, Phys. Lett B {\bf 646} (2007) 63.
\bibitem{page}D.N. Page, Phys. Rev. D, {\bf 13}(2) (1976) 198.
\bibitem{a9}S. Doplicher, G. Piacitelli,  L. Tomassini, S. Viaggiu, arXiv:1206.3067
\bibitem{NJ}E. T. Newman, E.T., Janis, A , J. Math. Phys. {\bf 6} (1965) 915.
\bibitem{z1}L. Accardi, F. Fidaleo, Rep. Math. Phys. {\bf 77} (2016) 153.
\bibitem{z2}F. Fidaleo, S. Viaggiu, Physica A {\bf 468} (2016) 677.
\bibitem{VVV}S. Viaggiu,  Int. J. Mod. Phys. D {\bf 9} (2006) 1441.
\end{thebibliography}
\end{document}